\documentclass[preprint2]{aastex}
\usepackage{natbib}

\shorttitle{EZ: AUTOMATIC REDSHIFT MEASUREMENT}
\shortauthors{Garilli et al.}

\begin{document}
\title{EZ: A TOOL FOR AUTOMATIC REDSHIFT MEASUREMENT}
\author{B.Garilli, M.Fumana, P.Franzetti, L.Paioro, M.Scodeggio}
\affil{IASF-Milano, INAF, Via Bassini 15, I-20133, Milano, Italy}
\author{O. Le F\`evre}
\affil{Laboratoire d'Astrophysique de Marseille, UMR 6110
CNRS-Universit\'e de Provence, 38 rue Frederic Joliot-Curie, F-13388
Marseille Cedex 13, France}
\author{S.Paltani}
\affil{ISDC, Geneva Observatory, University of Geneva, ch. d' \`Ecogia
  16, CH-1290 Versoix, Switzerland}
\author{R.Scaramella}
\affil{INAF, Osservatorio Astronomico di Roma, via Frascati 33, 0040
  Monte Porzio Catone (RM), Italy}
\begin{abstract}
We present EZ (Easy redshift), a tool we have developed within the
VVDS project to help in redshift measurement from otpical spectra. EZ has been designed
with large spectroscopic surveys in mind, and in its development particular care has
been given to the reliability of the results obtained in an automatic
and unsupervised mode. Nevertheless, the possibility of running it
interactively has been preserved, and a graphical user interface for
results inspection has been designed.
EZ has been successfully used within the VVDS project, as well as the
zCosmos one. In this paper we describe its architecture and the algorithms used, and
evaluate its performances both on simulated and real data. EZ is an open source program, freely downloadable from http://cosmos.iasf-milano.inaf.it/pandora.
\end{abstract}

\section{Introduction}
Thanks to larger telescopes and more powerful instruments, during the
last decade we have witnessed an explosion in the size of
spectroscopic surveys both in the nearby and in the more distant
Universe:  from the hundreds of objects of the
surveys carried out in the early nineties (ESP, \citet{ESP},CFRS, \citet{CFRS}) to the
thousands of galaxies of present days deep spectroscopic programs
(zCosmos, \citet{zcosmos_main}, VVDS, \citet{vvds_main,vvds_wide} ,
DEEP2, \citet{deep2}) and to the hundred thousands of more nearby surveys
(e.g. 2dF, \citet{2df}, SDSS, \citet{sdss7}, 6dF, \citet{6df}) or to the hundred thousand or million objects of
the just started programs (VIPERS, \citet{VIPERS}, WiggleZ,
\citet{WiggleZ}, BOSS, \citet{BOSS}) and the few millions of galaxies
of the future (e.g. EUCLID, \citet{euclid}).\\
Such an evolution in survey size requires an analogous step forward in
the way the data are reduced and analyzed, making compulsory the use of
reliable automatic tools.
Among the various tasks to be done when carrying out a redshift
survey, redshift
measurement is one of the most demanding in terms of human resources
and skills required. Historically, the {\it rvsao} \citep{rvsao} package within IRAF has
been the first widely used program which
has been adopted for redshift measurement. It makes use of emission
lines ({\it emsao} task)
and of the continuum shape ({\it xcsao} task), on the basis of a correlation of the
spectral continuum against galaxy or stellar templates, implementing
the algorithm originally proposed by \citet{tonryDavis}. Although very powerful,
it does not foresee to weigh spectral
features by an error spectrum. When the VIMOS-VLT Deep Survey (from
here on, VVDS) started, it was
immediately understood that this could be a major drawback in using
{\it rvsao} within that project: VIMOS Low Resolution Red (LR-Red) spectra, like those obtained within the VVDS
project, are affected by heavy fringing redwards of $\sim8000\mbox{\AA}$. As a
consequence, spectra show features which are purely due to fringing but
which can easily be mistaken with emission lines by any automatism, as
well as by humans, unless they are properly weighted. 
Furthermore, if both emission and absorption lines are to be
used, some scripting is required to couple together results from the
pure cross correlation algorithm and the emission lines search
one. Last, but not least, the IRAF implementation of {\it rvsao} did
not allow a direct interface with the VIPGI package used for VVDS Data
reduction \citep{vipgi}.\\
For all these reasons, we have decided to develop EZ (Easy redshift,
pronounced ``easy''),
a new tool which
inherits from the {\it rvsao} experience and the cross correlation
technique, but allows to be easily plugged within VIPGI and 
could also be used as a VO-compliant tool via SSAP \citep{SSAP} or other
communication protocols. 
From the user's point of view, the most important requirement was to
be able to run the program both in a completely unsupervised and in 
fully interactive mode, with a user friendly
graphical interface to inspect the results obtained using different
parameters or templates. Furthermore, it must be possible to use
templates which do not necessarily cover the full spectral range for
any redshift value. Finally, it has to be easily usable for different
projects, i.e. all parameters must be defined either through a project
dependent parameter file or via command line. EZ has been initially
based upon the prototypes  KBred (Scaramella 1999,
unpublished) and YAZ (Scaramella 2004, VVDS team Internal report), 
with which the first 10000 VVDS spectra were measured, and which
contained the correlation and fitting algorithm described below.
\\
In section \ref{concept} 
we illustrate the rationale of EZ. 
Section \ref{architecture} gives a short
illustration of the software architecture and the most important
algorithms used are illustrated in section 
\ref{algorithm}. Section \ref{dec_tree_verbose} describes
in full how a redshift can be measured by EZ and section
\ref{reliability} illustrates how a reliability flag is attached to the
best solution proposed. In sections
\ref{results_sim} and \ref{results_real}
 we evaluate the performances on 
simulated and real data respectively. Finally section \ref{gain}
shows how much time and efforts can be spared using a 
tool like EZ when carrying out a large redshift survey.

\section{EZ basic concepts}
\label{concept}
There can be several approaches to perform redshift measurement: from
the most simple ones (like emission line finding or cross
correlation) to more sophisticated ones, like bayesian approach or
principal component analysis. Within EZ, we have chosen to develop a
kind of {\it expert} system: the core of EZ is a decisional tree which
tries to mimick the decisional pattern followed by astronomers when
measuring redshifts from spectra. 
When an astronomer manually measures a redshift, his brain
almost unnoticeably performs several functions: he looks for the
existence of emission lines, and if there are some, looks whether they
match on a single redshift solution; he looks at the shape of the continuum, to
find out whether the object is an early type galaxy, and in that case
searches for a D4000 break and  Ca H and K absorption lines; discards
parts of the spectrum affected by heavy noise and spurious features,
like zero orders or sky subtraction residuals; possibly, looks at the
raw 2-dimensional spectrum to decide whether emission features are real
lines, or sky subtraction residuals.
The basic idea we have tried to implement in EZ is to allow the user to
combine the available functions in 
the most appropriate way for the data at hand, thus building new user
defined functions and methods. At the upmost level, a redshift measurement
{\it decisional tree} can be built, which mimicks the decisional
path followed by an astronomer to get to the measure of the redshift. 
Such decisional path can differ from data
set to data set, and for this reason several {\it
  decisional trees} can coexist in EZ, exactly as several
implementations ({\it methods}) of the same function can coexist.\\
EZ implements several {\it functions} to perform the different tasks
required: 
read data files and
templates, subtract the continuum from a spectrum, 
find emission lines, measure the redshift, 
measure spectral features.
Each of these functions, in turn, supports different {\it methods},
i.e. can use different algorithms to carry out that particular task: 
for example, fitting can use a chi-square
minimization or some more sophisticated algorithm, correlation can be
performed weighting data by a noise spectrum or without any weight,
data can be read from fits or ascii files.\\
Before coming to a detailed description, it is worth mentioning the
importance that spectral templates have during the redshift
measurement process. Whenever the solution relies on correlation or
fitting methods (i.e. is not based solely on emission lines matching), the reference templates used in the process should
reproduce as much as possible the spectra to be analyzed. It is not always
straightforward to have at hands templates for all possible galaxy types and
covering a wavelength range as large as to cover from the UV to the
NIR rest frame. Nor the same template library is necessarily the best suited for all
projects. EZ does not provide templates itself, but allows the user to
have different template suites and use those he thinks are more suitable for
that particular data set. Furthermore, it is possible
to specify a minimum range of overlap between the template
and the spectrum to be used. This feature eases a
bit the problem of providing templates covering an extremely large
wavelength range when dealing
with data sets sampling galaxies from the nearby universe up to $z>4$.\\

\section{Software architecture}
\label{architecture}
EZ is
implemented in Python in the form of a command-line interpreter. It
consists of a main Python class which imports other classes dedicated
to redshift estimation, service functionalities, file I/O, line
finding, line matching, and others. A complete description of the EZ
classes is provided in the downloadable user's manual.
In general, only the higher level methods are used (e.g. the
decisional tree) from within the EZ environment, but experienced users can directly use the Python
shell to import the various classes and combine lower level methods
in different ways, thus exploiting at full the flexibility of the
package. 
Even at the higher level, several sessions (i.e. instances of the same
classes) can coexist, and users can directly compare the results
obtained
using different methods or functions.\\

Algorithms are implemented in
Python, or in C for the most CPU intensive tasks: 
all algorithms implemented in C are called through a unique C
interface, so that all of them can easily be used from a Python
environment. 
All the classes are built in such a
way that they can be imported as modules
from any other Python code as well as from the Python shell. The main class
contains methods that return the best solution obtained as well as a 
"theoretical" spectrum for a given template
and a given redshift normalized to the input spectrum. Any other
Python code can simply import the main class and then directly handle the
proposed solution according to needs. \\

\begin{figure*}
\includegraphics[clip=true,scale=.70]{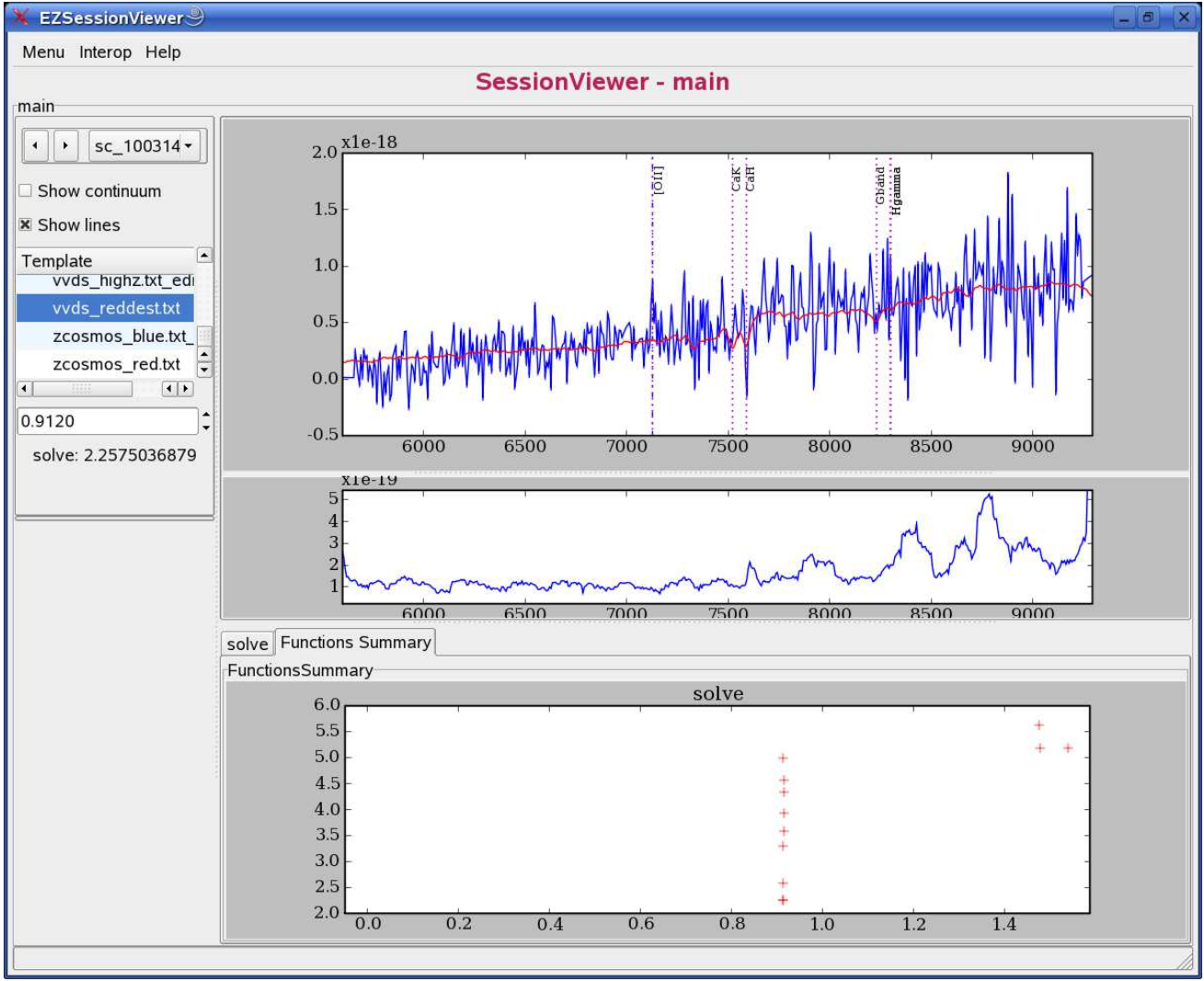}
\caption{EZ Graphical User Interface. In this figure we show a low
signal to noise spectrum with the purpose of demonstrating EZ
capabilities also on low quality data. Top panel: observed spectrum
  with best fitting template superimposed (red line). Most important
  emission or absorption lines are
  shown as vertical dotted red lines, emission lines found by EZ are marked
  with vertical blue dashed-dotted line. Middle panel: the noise
  spectrum. Bottom panel: the best reduced $\chi^{2}$ found for each template as a
  function of redshift. Left panel: top, pull down menu with the list of
  currently loaded spectra; middle, the list of available templates,
  the one currently shown is highlighted; bottom, the redshift of the currently
  displayed solution (default to the best solution found) and the
  corresponding reduced $\chi^{2}$}
\label{EZ_Gui}
\end{figure*}

While EZ is primarily developed as a command-line tool, a gtk based graphical
interface is available to the user. By design, it is "merely" an interface
to the command-line interpreter: it shows the spectrum and the
associated noise spectrum, allows superposition of the best fitting
solution, lists the other (less probable) solutions and allows to
overplot them onto the spectrum. An example of the Graphical User
Interface is shown in Fig. \ref{EZ_Gui}. In this figure we show a low
signal to noise spectrum with the purpose of demonstrating EZ
capabilities also on low quality data.\\

\section{Algorithms}
\label{algorithm}
\subsection{Emission line finding}\label{linefind}
One way to measure a redshift is to look for emission lines on to
which to
anchor a
solution. The basic concept of an emission line search
algorithm is to look for sharp peaks in the spectrum, as candidate
emission lines, and then see whether such peaks can be matched with a
single redshift solution. In presence of fringe patterns which
highly resemble emission lines, the main problem is to discard those peaks which are
very likely fake features, without loosing true lines. The implemented
procedure first
builds a {\it rough peak list} containing the position of all pixels
showing a flux above a user defined significance threshold. A gaussian is
then fitted to each of such positions, and only those peaks for which
the width of the fitting gaussian is within some user defined limits are retained: the
minimum and maximum width of a line depends on the resolution of the
spectrum, and as such are configurable. This check discards peaks due
to non or badly cleaned cosmic rays, which are too narrow to be a
real line. Among the remaining line candidates, a further check
is made whether the real peak flux is within a factor of two from the
fitted gaussian peak. This allows to discard most of the fake peaks
due to fringing which often have an irregular shape. Depending on
spectral resolution, partially resolved lines can also be poorly
represented by a gaussian. For this reason such check can be tailored or
altogether suppressed through user parameters.
Finally, the significance of the peaks is computed, subtracting a
local continuum and computing the ratio between the peak height and the
noise weighted local continuum: only peaks showing a significance above a pre
defined $\sigma$ (usually 4 or 5) are retained. As the noise is stronger in fringing
affected regions, spurious peaks with a gaussian shape get discarded
by this check. 
The lines thus identified are further subdivided into {\it strong}
  and {\it normal} lines, according to whether the peak value is higher than 
${\it strongcut}*\sigma$
where {\it strongcut} is a user defined parameter. 
{\it Strong} lines are treated differently within the decisional tree
  (see section \ref{dec_tree_verbose}).\\
Finally, the algorithm tries to match all or some of such peaks
to known combinations of emission lines at different redshifts. The
lines to be used for the matching are defined in a configuration
file, and can be changed at wish, according to the redshift range or
the type of object 
explored by the survey.\\

\subsection{Correlation}
Before applying the correlation both the
spectrum and the template are continuum subtracted so that what
drives the results are the local features (such as weak emission
lines, absorption lines, spectral breaks).
Each available spectral template is redshifted to a given redshift and
the correlation function is computed.
During the computation, each pixel can be optionally weighted by its
associated noise.
In the noise weighted approach, the correlation function has the form:

$$c(z)=\frac{\sum_{j \in \Lambda}\frac{(s_j-\bar{s})(t_j-\bar{t})}{n_j}}
{\sum_{j\in \Lambda}
\frac{\sqrt{\sum_{j\in \Lambda}{(s_j-\bar{s})^2}}\sqrt{\sum_{j\in
      \Lambda}{(t_j-\bar{t})^2}}}
{n_j}}, z\in \Theta$$


where $s_j$ and $n_j$ are the spectrum flux and noise at pixel
$j$ respectively;
$\bar{s}$ is the spectrum mean computed as

$\bar{s}=\frac{\sum_{j \in \Lambda}{{s_j}/{n_j}}}{\sum_{j \in \Lambda}{{1}/{n_j}}}$

$t_j(z)$ is the interpolated flux of the template at pixel $j$, once put at redshift z;
$\bar{t}$ is the template mean;
$\Lambda$ is the wavelength  range in use and 
$\Theta$ is the redshift range to explore. 
Each template does not necessarily
cover the full spectral range, in other words for different templates
and redshifts the {\it correlation  function} can be computed using a
different number of points. In order to compare
the results obtained for different templates, the value of the 
{\it correlation  function} is normalized to the number of points used
($N$).

$$c(z)_{red}=\frac{c(z)}{N}$$

Once the {\it correlation
  function} is produced for a particular template, 
the highest peak is returned as the {\it best
  correlation solution} for that template.

\subsection{Fitting}
The fitting of a spectrum against a template is performed using a
non-continuum subtracted spectrum, so that the results are affected
also by the overall shape of the underlying continuum. It uses a
standard least squared metric:
each spectral template available is redshifted to a given redshift and
the mean square deviation between spectrum and template
is computed as

$$\chi^{2}_{red}=\frac{\sum_{j\in\Lambda}\left(\frac{f_j-At_j}{\sigma_j}\right)^2}{N}, z\in\Theta$$

where 
$f_j$ and $\sigma_j$ are the spectrum flux and noise at pixel
$j$ respectively; 
$t_j(z)$ is the interpolated flux of the template at pixel $j$, once put at redshift z;
$\Lambda$ is wavelength  range in use and 
$\Theta$ is the redshift range to explore; 
$N$ is the number of data points used in the computation. Each template does not necessarily
cover the full spectral range, in other words for different templates
and redshifts the $\chi^{2}$ can be computed using a
different number of points. In order to compare
the results obtained for different templates, the
$\chi^{2}$  is divided by the number of data points used
The normalization constant $A$ is computed as:

$A(\Lambda, z)=\frac{\sum_{j\in\Lambda}\frac{f_j}{\sigma_j^2}}{\sum_{j\in\Lambda}
\frac{t_j}{\sigma_j^2}}$.

The minimum reduced $\chi^{2}$ for each template and redshift range is returned.

\subsection{The solve method}
\label{solve}
When no emission lines have been found in a spectrum, or when the
lines found do not
point to one single solution, redshift determination can be done using
a correlation, or a fitting procedure, as
described above. The {\it solve} function within EZ is meant to combine
cross correlation, fitting or any other elementary method
which can lead to a redshift solution. 
As all other functions, it can have different methods, i.e. 
the various elementary algorithms can be combined in different ways
according to wishes.\\
The current
implementation of the {\it solve} method foresees the use of cross
correlation first, and a
further fitting step. 
We have noticed that  
in the VVDS and zCosmos data sets, 
the highest peak of the correlation function as
described above is not
always the best solution, and this depends mainly on the fact that the
noise spectrum does not (negatively) weigh enough the spurious features
of the fringing. On the other hand, also the simple fitting does not
give satisfactory results in terms of picking up the correct redshift.
A sequential usage of the two methods, instead, has proven to be the one giving
the best results for our data. First a correlation of the spectrum
against each template is performed: as described above, this step
allows to properly take into account the local features of the
spectrum, like spectral breaks, absorption lines and weak emission
features. At the end of the correlation, for each template 
the {\it n} redshifts (where {\it n} is a user defined parameter) 
corresponding to the highest correlation
peaks are retained. At this point, we are faced with {\it n}x{\it m}
solutions (where {\it m} is the number of available templates).
To discriminate among them, we
use the fitting procedure: as fitting is performed using the spectrum without
subtracting the continuum, the overall spectrum shape plays a role
into getting a lower reduced $\chi^{2}$, and
can help to pin point the correct redshift. Finally, the
solution giving the minimum reduced $\chi^{2}$ is chosen. \\

\section{EZ decisional tree}
\label{dec_tree_verbose}
 The core of EZ is the ``decisional tree'', which tries to mimick the
 human decisional process applied during redshift measurement. This is where 
one defines the actions to be performed,
and in which sequence they
 must be carried out in order to measure a
 redshift. Each logical block of a decisional tree makes use of the
 lower level algorithms available. 
 Several decisional trees can coexist in EZ: each one acts
 differently in different situations, and this can be useful to
 better tune the measurement process according to the kind of data
 at hand. For example, in a survey where stars have been a priori
 removed, it may be useful not to check for M stars, thus reducing
 the degrees of freedom of EZ, and consequently the possibility that
 it takes a false track. As an example, we illustrate here the decisional tree
 we have implemented for the VVDS and zCosmos surveys. The input
 consists of one or several observed spectra with a noise
 spectrum associated to each of them. Following fig \ref{dec_tree}, the steps performed
 are as follows

\begin{enumerate}

\item\label {Mstarcheck} Check if the object is an M type star: the
  characteristical wavy shape of the spectra of this kind of stars can
  be recognized by
  looking if around the expected position a very large gaussian can be
  fitted, and if the spectrum is steadily becoming redder and redder
  with increasing wavelength. 

\item\label {Mstar}If such check is positive, in order to assign a best-fitting template to
  the object a fit is made
  using only M-star templates, and the redshift is set to zero.

\item\label{2_ELricerca} If the {\it M star check} has failed, search for emission lines in the spectrum; if
  lines are found go to step \ref{2_ELmatch}, otherwise go to step
  \ref{2_NoEL} 

\item\label{2_ELmatch} Search for a match between  lines. 
If one or more matches with at least 2 lines have been
  found go to step~\ref{2_3match}, otherwise go to step~\ref{2_stronglines}. 

\item\label{2_3match}  Fit each of 
  the possible solutions given by the matches above with emission lines templates. 
  Choose the solution giving the minimum reduced $\chi^{2}$.

\item\label{2_stronglines} Check if strong lines have been found, (see \ref{linefind}). If
  this is the case go to step~\ref{2_Yesstrong}, otherwise go to step
~\ref{2_NoEL}.	

\item\label{2_Yesstrong} Using only the redshifts satisfying matches
  with the one strong line, and only the emission line templates, compute the redshift using the
  algorithm defined by the {\it solve method} (\ref{solve})

\item\label{2_NoEL} Using all selected templates and the whole redshift
  range indicated by the user, compute the redshift using the
  algorithm defined by the {\it solve method} (\ref{solve})
  
\end{enumerate}

\section{Redshift reliability}
\label{reliability}
Complete automation of the redshift measurement process can be tricky
when spectra are noisy (as they always are at the faint limit of a
survey)
or in presence of artifacts such as fringing
correction residuals, so that  it
is by no means guaranteed, a priori, that the best solution proposed
by EZ is also a correct solution. 
For this reason we
have added the computation of an integer reliability flag which
summarizes the goodness of the solution proposed. Such a flag is
computed mimicking the  kind of logical reasoning applied by an
astronomer when trying to
evaluate if a redshift is reliable or not. The flow is as follows

\begin{itemize}
\item each template is described by a ``reliability'' file. 
  An illustration of such file for an Elliptical galaxy and for
  a StarBurst galaxy, as used with the VVDS and zCosmos projects, 
  is given in Table \ref{rel_file_table}: in the
  first column, the features which are tested are given, together with
  their wavelength (column 2). Each feature has a {\it weight}
  associated (column 5), according to the prominence the feature usually has in
  standard spectra. For example, in elliptical galaxies the D4000 break is
  the most prominent feature, thus its weight is higher than all other
  lines. On the other hand, if the feature is not found, then the
  redshift becomes suspect, thus a negative weight is associated in
  this case (column 6). Some features are commonly found together with
  other features (e.g., the [OIII] doublet, or the D4000 which usually comes
  with Ca H and Ca K absorption lines). The third column lists such
  {\it secondary} features, if applicable. If these features are found
  together with the main line, then the weight of the main line is
  increased to the value listed in column 4. Note that what is
  important is not the absolute value of the weight itself, but its
  relative value with respect to the other expected fetaures. The {\it
  reliability} files should be created according to user's needs and
  to the templates used.
\item the ``best fitting'' template is redshifted to the measured 
  ``best redshift'', and a correlation around each expected spectral feature is
  performed, taking into account the noise spectrum. The correlation
  value must be above a user defined threshold for the feature being considered as
  ``found''. In this step, observational constraints are taken
  into account: e.g. if one of the expected lines falls outside the
  observed spectrum (or too close to the border) it is ignored.
\item if the ``best solution'' is a star also the ``color'' is
  computed, as the difference between the
  mean value in the bluer and redder part of the spectrum, and used as
  a ``feature'' (blue color for earlier star types).
  If the star type is M, a dedicated
  algorithm searches for the characteristic ``wavy'' shape and uses
  them as features found (or not found).
\item if a feature has been found, then its correlation value is
  weighted according to the weight given to that feature in the reliability file for that
  template (column 4 or 5 in Table \ref{rel_file_table}). 
  Features listed in the reliability file, but not found,
  are negatively weighted (column 6 in Table \ref{rel_file_table}).
\item the weighted correlation values are summed up, and normalized by
  the number of lines which have been found, giving a {\it rate}
\item finally, the  {\it rate} is converted into a flag, according to
  the ratio of found features with respect to expected features and to the {\it rate} itself: 
  the higher the rate and the number of features found, the higher the
  flag. The conversion is made in such a way that EZ reliability
  flags resemble as much as possible the VVDS reliability flags
  as defined in \citet{vvds_main} in terms of confidence level.

\end{itemize}

Flags thus computed range from 0 (solution is not reliable) to 4 (highly
reliable solution).  In Table \ref{flag_table} we summarize the
confidence level of the different flags as defined in the VVDS
project, and the criteria used by EZ to assign each flag.

\section{Performances on simulated data sets}
\label{results_sim}
For any given spectrum, EZ gives in output the best
redshift and the best fitting template, while the output of the {\it
  reliability} process is the flag and the number of features
found. When evaluating performances, we should make a distinction between the
performances
of the redshift measurement methods, i.e.
the capability of finding the
correct redshift, and the performances of the {reliability} method,
i.e. the capability of assessing the goodness of the solution. \\
To perform such evaluations, we have carried out two different sets of tests: a first set,
using simulated spectra, and a second set, making use of real spectra
from the VVDS and the zCosmos survey.
\subsection{Simulated test set}
Using simulated spectra to evaluate performances of an algorithm has
the advantage that the answer is known beforehand, and thus
results can be evaluated with no error margin. On the other
hand, even if simulations are carried out as carefully as possible, such set can only give an
upper limit to the performances, as simulated spectra never take into
account all possible noise sources and defects existing on real data.\\
The procedure we used to simulate spectra begins with an input catalog of
objects: in this case we have used the Cosmos Mock Catalog (from
here on CMC), 
described in \citet{CMC}. In brief, CMC is a simulated catalog built directly from the
observed COSMOS \citep{scoville} catalog of \citet{capak} and the
COSMOS photometric-redshift 
catalog \citep{cosmo_photoz}. 
In CMC, a redshift and a spectrophotometric type 
are associated to each galaxy of the COSMOS catalog, using
a model fitting procedure of the photometric data. The resulting
catalog contains a mix of galaxy populations which by construction is 
representative of a real galaxy survey. Emission line fluxes are also
computed and magnitudes in a number of filters are made available. 
Details on the simulation of observed spectra will be given 
elsewhere (Franzetti et al., in preparation). To summarize the
procedure, a
rest frame spectrum, as can be obtained from galaxy model libraries,
has been associated to each galaxy type as provided in the catalog (elliptical,
early spiral, late spiral and starburst galaxy). In this step, line
broadening due to galaxy velocity dispersion has been neglected, as at
the resolution of the VVDS and zCosmos data this it is irrelevant. 
The monodimensional incident spectrum has been obtained by 
redshifting the rest frame spectrum and normalizing it so as
to give the
object apparent magnitude in the chosen selection band,
$I_{AB}$. 
The incident spectrum has then been
degraded
for the VIMOS efficiency curve, as can be obtained 
from the ESO Exposure Time
Calculator. In the noise calculation, we have taken into account the
Poissonian noise (from both sky and object), the flat fielding accuracy
and the electronic noise (for a detailed explanation of the different
contribution see e.g. \citet{newberry}). The sky spectrum used has also
been derived from ESO VIMOS exposure time calculator. 
We have compared our simulated
spectra with those which are obtained from 
ESO exposure time calculator, once the same exposure
time, galaxy type and apparent magnitude are used, and the results are
extremely similar both in terms of signal to noise as a function of
wavelength, and in terms of sky subtracted spectrum.
It is important to note that our simulated data set includes the
electronic noise, the poissonian noise, the flat fielding accuracy but
does not include the effect of fringing.
\\
We have simulated $\sim11000$ galaxy spectra, in the magnitude range
$17.5 \le I_{AB} \le 22.5$ using the same exposure time as for the VVDS wide
survey and the zCosmos bright survey (\citet{vvds_wide} and \citet{zcosmos_main}),
and the VIMOS LR red grism.
CMC does not contain stars, but we are
interested to check EZ performances also on different types of
stars, as the selection function of galaxies for  surveys never
completely succeeds in excluding stars on the basis of photometry. Thus we have simulated ~11000 stars, with magnitude
ranging from 17.5 to 22.5, of different spectral types. The Pickles
stellar library has been used for their model spectra.
On this simulated data set, we have run  EZ in totally blind
unsupervised mode and
obtained
for each object a redshift and a redshift flag. 
The template set we have used in EZ is the one obtained within the VVDS
project and built from the VVDS data themselves: it comprises two
templates for early type galaxies, one template for Sbc galaxies, one
for Scd galaxies and 3 for starburst galaxies, with different
intensities of emission lines and line ratios. These templates extend
from $\sim3500\AA$ to about $\sim8000\AA$ rest frame and are
particularly suited to
search for a redshift solution within the redshift range 0 to 2.0,
a range which well matches the redshift range of the
simulated data given the magnitude cut we have imposed, in spectra covering the observed range
of the VIMOS LR-Red grism.
The measured redshift has then been compared to the real input
value, and classified as correct when the measured
value is within $10^{-3}$ of the real value, 
this limit corresponding to the theoretical 
redshift error given by the grism resolution. \\
\subsection{Global performances on simulated spectra}
In Table \ref{simul_table} the results of such comparison are
summarized: for galaxies and stars separately, as well as for the whole
sample(which has comparable numbers of galaxies and stars), we give the success rate (computed as the ratio between correctly
retrieved redshifts and total number of objects), the number of correct
redshifts, and the total number of objects to which EZ has assigned
the given flag. Results for each object type are splitted per EZ redshift flag. \\
Table \ref{simul_table} shows that on simulated data, EZ is exceptionally
good at retrieving the correct redshift, the success rate being 
97\% on the whole sample. In spite of the large number of simulated
spectra, very few objects are classified with a reliability flag of 1
or 2, so that the average success rate for these flags has not the
same statistical significance as the other flags. \\
Going deeper in the analysis of results, we note that stars are practically always recognized
as such, even if the flag associated to their measurement is extremely
low in half the occurrences. This is intrinsic to the way we
associate flags: for types younger than K, 
stellar spectra are poor in spectral features
in the wavelength range explored. At the resolution we used,
only NaD and H$\alpha$ absorption lines are clearly visible. Furthermore,
NaD falls very near to a strong sky line, and it is often not detected
due to the higher noise. Thus only one out of two features is
clearly detectable. This explains the frequency of zero flag objects in
this category. However, only 191 galaxies ($<2\%$) are erroneously mistaken for
stars, while among the 10762 objects classified as galaxies, only 19 were actually stars. \\

\subsection{Dependence on magnitude and redshift}
\begin{figure}
\plotone{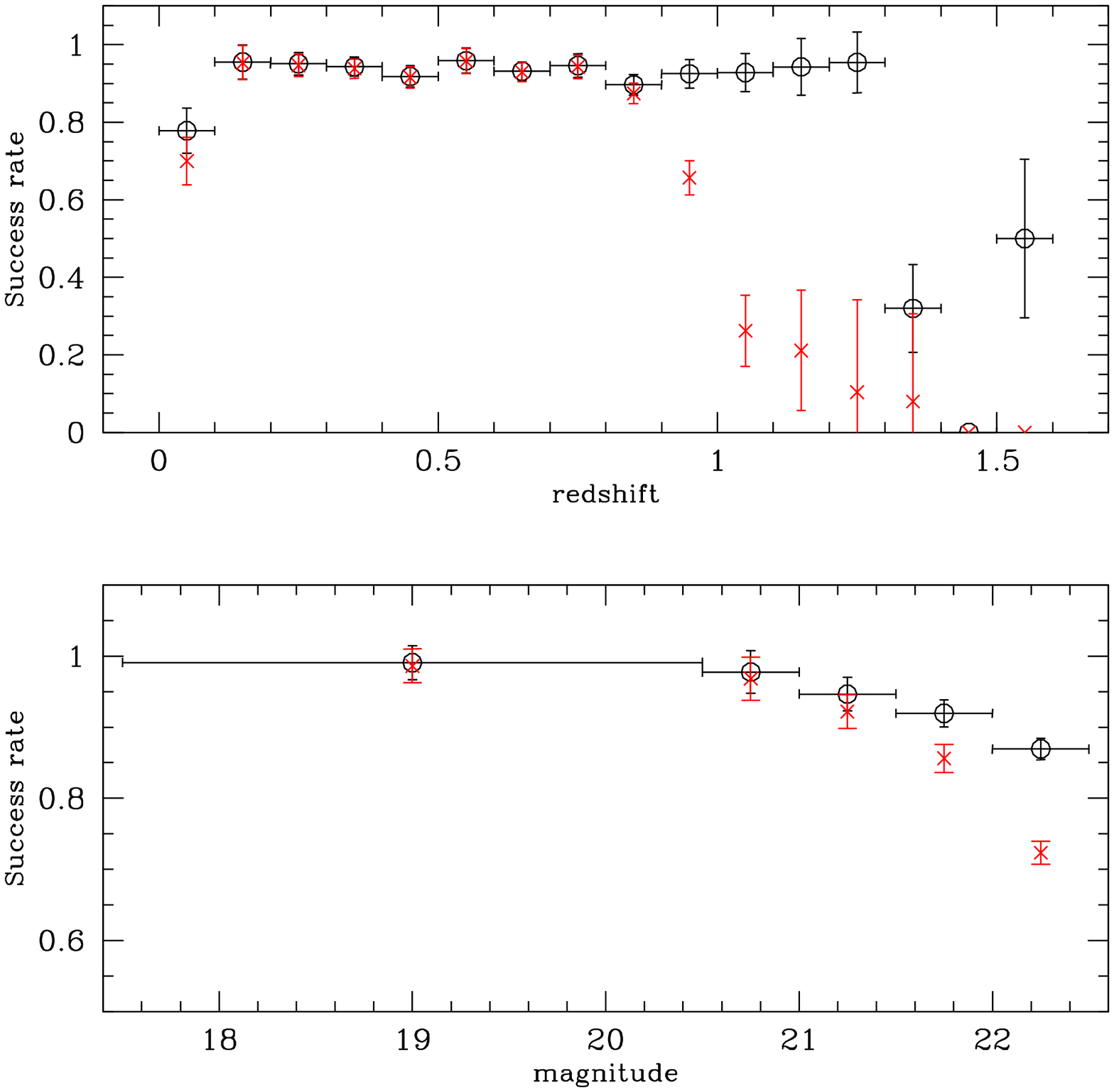}
\caption{EZ success rate as a function of magnitude (lower panel) and
  of redshift (upper panel), considering flags 2 and above (black
  open circles) and only flags 3 and 4 (red crosses). Horizontal error bars
  indicate the bin width, vertical error bars are poissonian errors}
\label{simul_perf}
\end{figure}

We have shown that on the simulated test set, the global success rate
is very high (97\%). Still we expect it to show a trend with object
magnitude, which can be considered a proxy for signal to noise ratio
for a fixed exposure time. This is shown in figure \ref{simul_perf},
bottom panel, where we plot the success rate obtained as a function of
magnitude. 
Black circles show the success rate obtained
considering all flags above 1, while red crosses
show the success rate obtained considering {\it very secure} redshifts
(flags 3
or 4) only. Even using all flags above 1 ({\it secure} redshifts), it is evident that
the fainter objects have an increasingly lower success rate, even if
it stays always well above $80\%$.\\
The top panel of figure \ref{simul_perf} shows the success rate as a
function of the true redshift of the object. The drop in success rate
for $z>1.4$ for {\it secure} redshift flags is not very surprising:
between z=1.4 and z=1.5, both the [OII] line and the D4000 break start
to fall out of the observed wavelength range, and the redshift can be secured
only for very few galaxies. In other words, we are entering in the
redshift desert regime, were observations in the red visible part of
the spectrum are known to be inefficient.
Possibly more surprising, at first glance, can be the drop in success
rate observed above z=0.9 when only {\it very secure} redshifts are
considered. This is mainly due to the way we compute the reliability
of the redshift: at z=0.9, $[OIII]_a$ and H$\beta$ lines progressively fall out
of the observed wavelength range, so that the only strong and easily
detectable line for late objects types remains [OII]. For this reason,
many objects get a flag 9 (only one strong line). On the other hand,
going higher in redshift objects become fainter, and their continuum
gets noisier: therefore also early type objects are more difficult to
measure and to get a {\it very secure} flag is even harder.

\section{Performances on real data sets}
\label{results_real}
\begin{figure}
\plotone{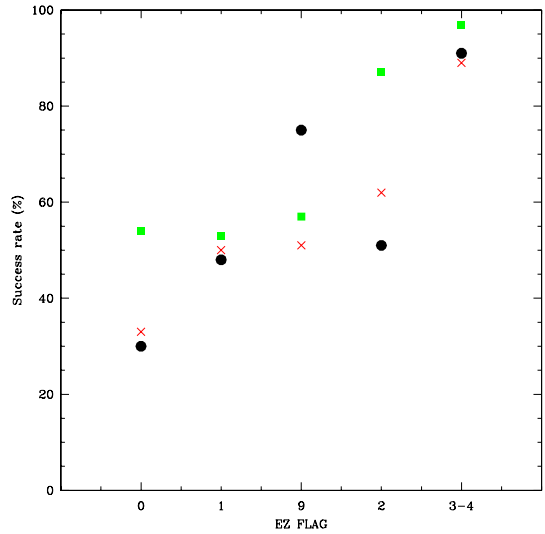}
\caption{EZ success rate as a function of reliability flag. 
Black dots for VVDS Deep sample, red
  crosses for VVDS Wide sample , green squares for zCosmos bright sample}
\label{performance}
\end{figure}

We have tested EZ performances on the three data sets of the VVDS Deep
survey \citep{vvds_main}, the VVDS Wide survey \citep{vvds_wide} and
the zCosmos survey \citep{zcosmos_main}. The three data sets have
different characteristics, and are complementary to evaluate EZ
performances: the VVDS Deep sample is cut to a deeper apparent magnitude
limit ($I_{AB}\le 24.0$), thus its redshift distribution extends beyond z=1.5 with
significant numbers. On the contrary, the VVDS Wide is cut at a brighter
limit ($I_{AB}\le 22.5$), and has a stronger star contamination. The zCosmos sample,
finally, has the same depth as the VVDS Wide sample, but it has been
observed with a higher resolution grism and stars have been a priori 
discarded on the basis of their morphologies and spectral energy
distributions: the criteria used were intentionally quite conservative
and a small fraction of stars is expected in the sample
\citep{zcosmos_main}. 
All redshifts for these
three samples 
have been manually measured by two different persons (also using beta
versions of EZ in
interactive way), discrepant
measures have been reconciled, and a flag has been assigned
to each redshift, corresponding to a confidence level as described 
in table \ref{flag_table}, column 2, a long and time consuming
procedure which has requested big efforts on the part of the people
involved. 
{\it A posteriori}, we have run EZ in blind and unsupervised mode on
the data sets, using only objects with a measured redshift between 0
and 2.0.
By comparing the results found by EZ with those published
by the VVDS and zCosmos consortia, we can evaluate EZ performances
on real data. We expect such performances to be worse than what
obtained on simulated data sets, as now all possible error sources are
present, last but not least the fringing above $\sim 8000 \mbox{\AA}$.
\\
Table~\ref{perfTable} and Figure \ref{performance} show the success rate of the EZ blind
measurement in the three different samples. As expected, the global 
performances are not as exceptional as for the simulated data, ranging from $\sim68\%$ to $\sim76\%$ in the 
best case of the zCosmos data. However, a straight comparison is not 
totally fair: as declared in the surveys themselves and reported in Table
\ref{flag_table}, measured redshifts are never claimed to be 100\% 
correct. The last column of Table \ref{perfTable} reports the 
weighted success rate:  for each EZ flag, the non concordant redshifts 
have been weighted in number according to the humanly given flag, so 
that discordant objects having a human flag 1 count by half, if they 
have human flag 2 or 9 they are weighted 75\%, if they have human flag
3 and 4 they are weighted 90\% and 95\% respectively. Using such weights, 
the success rate obviously increases,as the 
uncertainty of the humanly measured redshifts is properly taken 
into account.
\\
Still, the success rate remains lower than what we had for simulated
data, especially in the VVDS samples. The main reason for this are the
fringing features, which if not properly weighted are easily
mistaken for real features by any automatic procedure. 
As explained in the previous
section, in the simulated data set we have not included the effect of
fringing, and for this reason simulated spectra are much less noisy redwards of
$8000\AA$ and EZ is not fooled by spurious emission features. The
noise spectrum associated to real data should help in weighting considerably
less such features, but when fringing features are particularly
strong, the noise spectrum, as currently computed, 
is not enough to completely neutralize their effect on
the emission line finding algorithm. As a test for this hypothesis, we
have run EZ on these same data sets without using the noise spectrum and the
resulting success rate is considerably lower. 
In the zCosmos sample, thanks to the larger
dithering width  the fringing pattern is better removed from the data
and the associated noise spectrum better allows to take into account
any residual. This is one of the reasons why in the zCosmos 
sample the success rate for the
most secure EZ flags is the same as for simulated data. A
second important effect is the presence of the zero order in the
extracted spectra: if this is not removed, it is mistaken for an
emission line and the redshift is obviously wrong. Again, zCosmos data
are less affected because the multiplexing of those observations is
much lower than for the VVDS data. Finally, also the observational 
conditions have an impact on
the data quality and the redshift measurement: clouds and bad seeing
diminish the S/N
ratio, a lower
exposure time as well as a too high airmass decrease the signal to
noise, bad centering of the object in the slit increases slit losses. All
these factors contribute to lower the global redshift measurement
success rate with respect to the idealized case of simulations.\\
Looking deeper at the concordant measurements, it can be noted that the
flag assigned by EZ is not always identical to that assigned by the
astronomers. This is summarized in Table \ref{real_flag_table}, where
we compare EZ flag with the {\it human} flag for concordant
measurements. In this table, we use only measurements obtained for
galaxies, as we have already shown that when the object is a star the
flagging system we have implemented gives low reliability values. We
have grouped results for flags 2 and 9, because experience has taught that
astronomers' ideas on the two flags differ from person to person: when
only one emission line is clearly visible, some people look at the
continuum slope and assign a flag 2, others don't. 
Table \ref{real_flag_table} shows that the similarity of the two
flagging systems is confined to the highest confidence bin: the vast
majority of the most reliable measurements by EZ are considered {\it
  very secure} also by astronomers. Going to lower EZ flags, the
astronomer's flag is usually higher than what judged by the 
automatic algorithm.
This effect is not unexpected, since the EZ flagging system has been set up
to be as reliable as possible, at the expenses of being too
pessimistic in many cases. We must also consider that the criteria
used by EZ and by an astronomer are not exactly the same: astronomers
only grossly take into account the correlation with the continuum
slope, while in absence of emission lines this is very important in
an automatic tool. Also the lines to be searched must be very significant
for EZ to take them into account, and the significance is weighted
with the corresponding noise spectrum, while an eye measurement is
more elastic in judgment, also because of the possibility of
double-checking
1D with 2D spectra. 
Finally, the astronomer's judgment is
subjective and changes not only from person to person, but also with
the tiredness of the person performing measurements. An automatic
evaluation, on the contrary, although more pessimistic, is always based on the
same criteria. 
\\
As reported in the papers presenting the surveys (\cite{vvds_main}, \cite{vvds_wide} and
\cite{zcosmos_main}), 
a redshift has not been measured for all objects.
These measurement failures are indicated in the catalogs
with a conventional {\it human} flag of zero and no redshift
indication, thus they are not included in table
\ref{perfTable}. Nevertheless, it is interesting to see which flag EZ
gives to these critical data. In all the three samples we have
considered, more than 80\% of the spectra for which 
astronomers did not measure a redshift get a flag 0 or 1 from EZ, a clear indication of a highly
unreliable solution. In the remaining few cases, EZ gets 
fooled by fake features, indicating that the noise spectrum is not
accurate enough.
\\ 
\section{Application on large extragalactic redshift surveys}
\label{gain}
The original purpose for which EZ has been developed is to ease and
speed up the long and painful phase of measuring redshifts in large
surveys, reducing human intervention as far as possible while keeping
the redshift measurement success rate
acceptable for the scientific purpose of the project. 
On the basis of the results shown in Table \ref{simul_table},
and keeping in mind such purpose, we can explore which would be
the limitations of adopting straightforwardly EZ results in measuring
redshifts in a generic redshift survey performed in optimal conditions, as
simulated data are, which explores the redshift range we have
considered here, i.e. $0<z<2$.\\
We define as completeness the number of 
measured redshifts over the total number of spectra, while purity is the number
of correctly recovered redshifts with respect to the measured ones. The results are given in table
\ref{example}: 
accepting blindly EZ results for all objects with a measured
redshift $>0$, irrespective of flag, the
survey would be $>98\%$ complete, with $<6\%$ of wrong
redshifts. According to the degree of completeness which can be
sacrificed to purity, it is possible to retain only galaxies with
very high reliability flag (3 or 4): as from table \ref{example}, in this case the
resulting galaxy sample would be slightly less than $90\%$ complete,
but its purity would be extremely high ($>95\%$). If we apply these
percentages to next generation surveys, where the number of spectra
foreseen is of the order of few hundred thousands, the {\it gain}
(meant as number of spectra for which human measurement can be
avoided) 
in using a
reliable automated redshift measurement tool is evident.\\
One may argue that real surveys are not always carried out in
optimal conditions, real data are usually more difficult to treat
than simulated data and as such the percentages given in table
\ref{example} are only upper limits. In section \ref{results_real}, 
we have shown 
 that EZ is extremely reliable
to find the correct solution for spectra classified as {\it very secure} 
(the success
rate being above 90\% for EZ flag 3 and 4) even on real data, affected
by different sources of noise. Using these encouraging
results,
we can extend the previous considerations to more realistic sets and
estimate what would be the {\it gain}, in terms of number of spectra
which can skip the human check, if we would fully trust the redshifts to which EZ assigns
a flag 3 or 4.
This is summarized in table \ref{gain_table}, where for each of the
real samples we have analyzed we give the total number of objects observed, the number
of objects for which EZ measured a highly reliable redshift, the
purity of this {\it highly reliable} sample and the {\it gain} meant
as the fraction of measurements which could be avoided. Such gain ranges
from 40 to more than 50\%, and it translates into a factor of almost 2
sparing of human effort and
time.  Given the difficult nature of the data
used for our tests (see the discussion on fringing effects in the
previous section) such a number can be considered as a lower limit to
the effective gain obtainable by using EZ in a large extragalactic
redshift survey.

\section{Summary}
We have presented, EZ, an automated tool devoted to redshift
measurement. The concept at the basis of the tool is the {\it
  decisional tree}, i.e. the sequence of operations to be performed to
obtain a redshift. Such sequence of operations can be customized
according to the kind of sample at hand: e.g. by discarding any check
on stars, or using only emission lines without performing the correlations
on the continua in the case of spectra obtained with an instrument
operating in slitless mode. \\
EZ can be used both interactively, with the help of a graphical user
interface, or totally blindly in unsupervised mode. It is developed in
Python, with the bulk of computations performed in C to increase its
computational speed. Its Python classes can be directly imported in
any other Python based program, thus making it fully embeddable in any
application.\\
Within EZ we have developed a method to assign a reliability flag on
the measurement obtained, in order to mimic the kind of reasoning done by
astronomers when assessing the goodness of the solution found.
The implemented flagging system,
though, is rather conservative.\\
We have tested EZ on VIMOS-like simulated data, and have shown that
its performances are exceptionally good, the redshift measurement
success rate being above 95\%. \\
We have blindly applied EZ to  the VVDS-Deep,
VVDS-Wide and zCosmos bright samples, and have demonstrated how this
tool behaves very well also on real data. The success rate obtained is
around 70\%, and rises above 90\% for redshifts classified 
as {\it very secure} by astronomers.\\
Finally, we have shown that the adoption of a similar tool can save
from a minimum of 50\% to a maximum of 95\% of the redshift
measurement load, according to the quality of the data and to the
degree of completeness and purity of results one is willing to
sacrifice in favor of a  fast output. \\
EZ is now being routinely and blindly run as part of the reduction 
process within the VIPERS survey, while a customized
version has been set up to perform simulations of the E-NIS
spectrograph \citep{euclid}.\\
EZ is an open source program, freely downloadable from http://cosmos.iasf-milano.inaf.it/pandora
\begin{acknowledgements}
We wish to thank the whole VVDS and zCosmos collaboration for their testing of the earlier versions of EZ and their helpful feedback. A special thank to Dario Maccagni, for his invaluable help
in preparing the manuscript. This work has been fully supported by INAF, through funding by the Project Department via the Information Systems unit. 
\end{acknowledgements}

\bibliographystyle{aa}
\bibliography{Garilli_revisedVersion}

\clearpage

\begin{deluxetable}{cccccc}
\tablewidth{0pt}
\tablecaption{Examples of reliability weighting scheme
\label{rel_file_table}}
\tablecolumns{6}
\tablehead{
\colhead
{Feature} & {wavelength} & {secondary} & {weight if}
 & {weight if secondary } & {weight if} \\
{name} & {} & {feature} & 
  {all found} & {not found } & {
  not found} 
} 
\startdata
\cutinhead{StarBurst galaxy}
$[$OII$]$   & 3727.5 &           & 10 &     &-20 \\
H$\beta$  & 4861.3 & H$\gamma$ & 25 & 10  & -20 \\
$[$OIII$]_{a}$ & 5006.8 & $[$OIII$]_{b}$ & 30 & 10  & -20 \\	  	
H$\alpha$ & 6562.8 & SII       & 25 & 10  & -20 \\
\cutinhead{Elliptical galaxy}
D4000 & 4000   & CaH CaK  &   50 & 20 &  -10   \\
G$_{band}$ & 4304.4 &          &   10 &    &  0   \\
H$\gamma$    & 4340.4 &          &   10 &    &  0   \\
MgI   & 5175.4 &          &   10 &    &  0   \\
Ca+Fe & 5269.0 &          &   10 &    &  0   \\
NaD   & 5892.5 &          &   10 &    &  0   \\
\enddata
\tablecomments{The weighting scheme presented is the one adopted
  for the VVDS and zCosmos bright surveys}
\end{deluxetable}

\begin{deluxetable}{cccc}
\tablewidth{0pt}
\tablecaption{Significance of VVDS flags and criteria for EZ flags
\label{flag_table}}
\tablecolumns{4}
\tablehead{
\colhead
{Flag} & {VVDS} & {EZ lines} & {rate} \\
{} & {confidence}}
\startdata
4 & $>$95\% & $N_{det}/N_{exp}>0.5$ & high\\
3 & 90\%  & $N_{det}/N_{exp}>0.5$ & low \\
2 & 75\% & $N_{det}/N_{exp}<0.5$ & high\\
9 & 75\% & $N_{det}=1$  & high\\
1 & 50\% & $N_{det}/N_{exp}<0.5$ & low \\
0 & 0\%& $N_{det}=0$ & low\\
\enddata
\tablecomments{$N_{exp}$ and $N_{det}$ indicate the number of expected
  and detected spectral features respectively. See section
  \ref{reliability} for a detailed explanation.}
\end{deluxetable}

\begin{deluxetable}{cccc}
\tablecolumns{4}
\tablewidth{0pc}
\tablecaption{EZ performances on simulated data
\label{simul_table}}
\tablehead{
\colhead
{EZ flag} & {SR} & {Ncorrect} & {Ntotal}
}
\startdata
\cutinhead{Galaxies}
any& 92\% &10092 &10934 \\
3-4& 95\% & 9307 &9709  \\
 2 & 35\% &   14 &40	  \\
 9 & 98\% &  708 &717	  \\
 1 & 29\% &   43 &144	  \\
 0 &  6\% &   20 &324   \\
\cutinhead{Stars}
any& 99\% &8103 &8122\\
3-4& 99\% &4320 &4321\\
 2 & 99\% & 129 &130	\\
 1 &100\% &   3 &3	\\
 0 & 99\% &3651 &3668\\
\cutinhead{Galaxies and stars}
any& 97\% &18196 &21955 \\
3-4& 97\% &13627 &14030 \\
 2 & 91\% &  143 &170	  \\
 9 & 98\% &  708 &717	  \\
 1 & 42\% &   46 &147	  \\
 0 & 93\% & 3671 &3992  \\
\enddata
\tablecomments{
The success rate (SR) is defined
as the percentage of simulated spectra (Ntotal) that are assigned
a correct redshift (Ncorrect).  These are tabulated as a function
of the EZ flag described in the text.
}
\end{deluxetable}

\begin{deluxetable}{ccccc}
\tablecolumns{4}
\tablewidth{0pc}
\tablecaption{EZ performances on real data
\label{perfTable}}
\tablehead{
\colhead
{EZ flag} & {SR} & {Ncorrect} & {Ntotal} & {SR weighted}
}
\startdata
\cutinhead{VVDS Deep}
any & 68\% & 5833 & 8514 &  76\% \\
3-4 & 91\% & 3845  &  4191 & 94\%\\
2 & 51\% &  220 & 427 & 61\% \\
9 & 75\% &  730 & 962 & 82\% \\
1 & 48\% &  372 & 762 & 58\% \\
0 & 30\% &  666 & 2172 & 40\% \\
\cutinhead{VVDS Wide}
any & 70\% & 12242 & 17436 & 78\% \\
3-4 & 89\% & 9091 &  10126 & 93\% \\
2 & 62\% &  810 & 1288 & 72\% \\
9 & 51\% &  263 &  509 &  59\%  \\
1 & 50\% &  698 &  1381 & 56\% \\
0 & 33\% &  1381&  4132 & 43\% \\
\cutinhead{zCosmos Bright}
any & 76\% & 6403 &  8404  &  80\% \\
3-4 & 97\% &  3735 &  3822  & 98\% \\
2 & 87\% &  445 &   510  & 90\% \\
9 & 57\% &  194 &   336  & 62\% \\
1 & 53\% &  677 &   1260  & 59\% \\
0 & 54\% &  1352 &   2476  & 62\% \\
\enddata
\end{deluxetable}

\begin{deluxetable}{cccc}
\tablecolumns{4}
\tablewidth{0pc}
\tablecaption{EZ and humanly assigned flags comparison
\label{real_flag_table}}
\tablehead{
\colhead
{} & \multicolumn{3}{c}{human flag}\\
\cline{2-4}
{EZ flag} & {3-4} & {2-9} & {1} \\
}
\startdata
\cutinhead{VVDS Deep}
3-4 & 84 \%& 15 \%& 1 \%  \\ 
2-9 & 42 \%& 54 \%& 4 \%  \\ 
1 & 55 \%& 41 \%& 4 \%  \\ 
0 & 39 \%& 52 \%& 9 \% \\ 
\cutinhead{VVDS Wide}
3-4 & 72 \%& 24 \%& 3\%  \\
2-9 & 26 \%& 61 \%& 12 \%  \\
1 & 39 \%& 52 \%& 9 \% \\
0 & 24 \%& 56 \%& 20 \% \\
\cutinhead{zCosmos Bright}
3-4 & 89 \%& 9\%& 2 \% \\
2-9 & 58 \%& 36 \%& 6 \%\\
1 & 74 \%& 20 \%& 5 \%\\
0 & 40 \%& 44 \%& 15 \%\\
\enddata
\end{deluxetable}

\begin{deluxetable}{cccccc}
\tablecolumns{6}
\tablewidth{0pc}
\tablecaption{Completeness and purity of the simulated galaxy sample
\label{example}}
\tablehead{
\colhead
{flag} & {Ninput} & {Nmeasured} & {Ncorrect} & {Completeness} & {Purity}} 
\startdata
any      & 10934  & 10762     & 10092    & 98.4\%  & 93.8\% \\
3-4      & 10934  & 9709      & 9307     & 88.8\%  & 95.8\% \\
2;3;4;9  & 10934  & 10466     & 10029    & 95.7\%  & 91.7\% \\
\enddata
\tablecomments{Completeness is
  defined as the number of 
measured redshifts over the total number of input spectra, purity is
  defined the fraction of correctly recovered redshifts with respect
  to the measured ones}
\end{deluxetable}

\begin{deluxetable}{ccccc}
\tablecolumns{5}
\tablewidth{0pc}
\tablecaption{Purity of the real data sample
\label{gain_table}}
\tablehead{
\colhead
{sample} & {Ntotal} & {EZ flag 3-4 } & {purity} & {gain}
}
\startdata
VVDS-Deep & 9742 & 3845 & 91\% & 40\%\\
VVDS Wide & 18984 & 10126 & 89\% & 53\%\\
zCosmos Bright & 8404 & 3822 & 97\% & 45\%\\
\enddata
\tablecomments{Gain is defined as the percentage of spectra for which
  human measurement can be spared}
\end{deluxetable}
\clearpage

\begin{figure}
\plotone{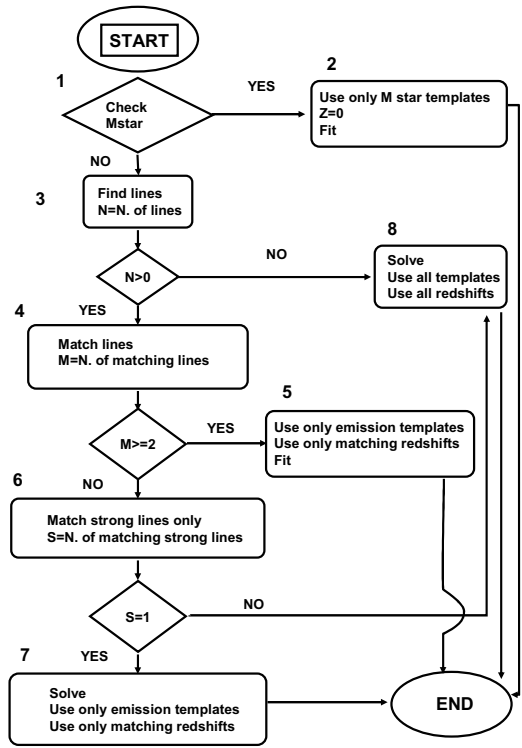}
\caption{Example of EZ decisional tree}
\label{dec_tree}
\end{figure}

\end{document}